\documentclass[nohyperref]{article}

\usepackage{microtype}
\usepackage{graphicx}
\usepackage{subfigure}
\usepackage{booktabs} %

\usepackage{hyperref}
\usepackage[accepted]{wfvml2023}

\usepackage{amsmath}
\usepackage{amssymb}
\usepackage{mathtools}
\usepackage{amsthm}
\usepackage{doi}

\usepackage[capitalize,noabbrev]{cleveref}

\theoremstyle{plain}
\newtheorem{theorem}{Theorem}[section]

\theoremstyle{definition}
\newtheorem{definition}[theorem]{Definition}
\newtheorem{assumption}[theorem]{Assumption}
\theoremstyle{remark}

\usepackage[textsize=tiny]{todonotes}

\newcommand{\R}{\mathbb{R}}
\newcommand{\N}{\mathbb{N}}

\newcommand{\IR}{\mathbb{IR}}

\newcommand{\calA}{\mathcal{A}}

\newcommand{\calR}{\mathcal{R}}

\newcommand{\calW}{\mathcal{W}}
\newcommand{\calX}{\mathcal{X}}

\newcommand{\sfF}{\mathsf{F}}

\newcommand{\bfu}{\mathbf{u}}

\newcommand{\bfw}{\mathbf{w}}

\newcommand{\ul}[1]{\underline{#1}}
\newcommand{\ula}{\ul{a}}
\newcommand{\ulb}{\ul{b}}
\newcommand{\ulc}{\ul{c}}
\newcommand{\uld}{\ul{d}}

\newcommand{\ulf}{\ul{f}}

\newcommand{\uls}{\ul{s}}

\newcommand{\ulx}{\ul{x}}

\newcommand{\ulC}{\ul{C}}

\newcommand{\ulN}{\ul{N}}

\newcommand{\ol}[1]{\overline{#1}}
\newcommand{\ola}{\ol{a}}
\newcommand{\olb}{\ol{b}}
\newcommand{\olc}{\ol{c}}
\newcommand{\old}{\ol{d}}

\newcommand{\olf}{\ol{f}}

\newcommand{\ols}{\ol{s}}

\newcommand{\olx}{\ol{x}}

\newcommand{\olC}{\ol{C}}

\newcommand{\olN}{\ol{N}}

\wfvmltitlerunning{Fast Interval Arithmetic in \texttt{numpy} for Formal Verification of Neural Network Controlled Systems}

\definecolor{tab:blue}{HTML}{1f77b4}
\definecolor{tab:orange}{HTML}{ff7f0e}
\definecolor{tab:green}{HTML}{2ca02c}
\definecolor{tab:red}{HTML}{d62728}
\definecolor{tab:purple}{HTML}{9467bd}
\definecolor{tab:brown}{HTML}{8c564b}
\definecolor{tab:pink}{HTML}{e377c2}
\definecolor{tab:gray}{HTML}{7f7f7f}
\definecolor{tab:olive}{HTML}{bcbd22}
\definecolor{tab:cyan}{HTML}{17becf}

\begin{document}

\twocolumn[
\wfvmltitle{A Toolbox for Fast Interval Arithmetic in \texttt{numpy} with an Application to Formal Verification of Neural Network Controlled Systems}

\wfvmlsetsymbol{equal}{*}

\begin{wfvmlauthorlist}
\wfvmlauthor{Akash Harapanahalli}{gtece}
\wfvmlauthor{Saber Jafarpour}{gtece}
\wfvmlauthor{Samuel Coogan}{gtece}
\end{wfvmlauthorlist}

\wfvmlaffiliation{gtece}{School of Electrical and Computer Engineering, Georgia Institute of Technology, Atlanta, USA}

\wfvmlcorrespondingauthor{Akash Harapanahalli}{aharapan@gatech.edu}
\wfvmlkeywords{Machine Learning, Formal Verification}

\vskip 0.3in
]

\printAffiliationsAndNotice{}  %

\begin{abstract}
In this paper, we present a toolbox for interval analysis in \texttt{numpy}, with an application to formal verification of neural network controlled systems.
Using the notion of natural inclusion functions, we systematically construct interval bounds for a general class of mappings.
The toolbox offers efficient computation of natural inclusion functions using compiled C code, as well as a familiar interface in \texttt{numpy} with its canonical features, such as $n$-dimensional arrays, matrix/vector operations, and vectorization.
We then use this toolbox in formal verification of dynamical systems with neural network controllers, through the composition of their inclusion functions.
\end{abstract}

\section{Introduction}
\label{sec:introduction}

Interval analysis is a classical field that provides a computationally efficient approach for propagating errors by computing function bounds~\cite{LJ-MK-OD-EW:01}. It has been successfully used for floating point error bounding in numerical and scientific analysis~\cite{intarith}. Interval bounds are often used for dynamical systems: (i) in reachability analysis, using methods such as Differential Inequalities~\cite{JKS-PIB:13} and Mixed Monotonicity~\cite{PJM-AD-MA:19,MA-MD-SC:21}; (ii) for invariant set computation~\cite{MA-SC:20}.
Recently, interval analysis has been increasingly used for the verification of learning algorithms: (i) standalone neural network verification approaches such as Interval Bound Propogation~\cite{SG-etal:19}; (ii) in-the-loop neural network control system verification approaches including simulation-guided approaches~\cite{WX-HDT-XY-TTJ:21}, POLAR~\cite{CH-JF-XC-WL-QZ:22} and ReachMM~\cite{SJ-AH-SC:23}.

Since all of these techniques use a similar suite of tools, there is value in creating an efficient, user-friendly toolbox for general interval analysis.
Python has become the standard for the learning community, and as such there are several existing tools for interval arithmetic. However, they come with key drawbacks:
\texttt{pyinterval} does not natively support interval vectors and matrices, and \texttt{portion} supports lists of intervals, but not matrix and vector operations.

\vspace{-1em}
\paragraph*{Contributions}
In this paper, we introduce a novel interval analysis framework called \texttt{npinterval}\footnote{The most recent code for \texttt{npinterval} can be viewed at \url{https://github.com/gtfactslab/npinterval}; to reproduce the figures in this paper, see \url{https://github.com/gtfactslab/Harapanahalli_WFVML2023}.}, implemented in \texttt{numpy}~\cite{numpy}, the computational backbone of most scientific Python packages.
This framework is built upon the notion of inclusion functions, which provide interval bounds on the output of a given function. We first define tight inclusion functions for several elementary functions, then use Theorem~\ref{thm:nif} to build natural inclusion functions for a more general class of composed functions.
The proposed package extends the prominent benefits of \texttt{numpy} directly to interval analysis, including its efficiency with compiled C implementations, versatility with $n$-dimensional arrays, matrix/vector operations, vectorization, and its familiar user interface.
We then demonstrate its utility through an application in formal verification of neural network controlled systems, by composing CROWN~\cite{HZ-etal:18}, a state-of-the-art neural network verification method with a natural inclusion functions of the system in Theorem~\ref{thm:clsysreach}. The proofs of all the Theorems are presented in Appendix~\ref{sec:proofs}.
\vspace{-2em}
\paragraph*{Notation}
We denote the standard partial order on $\R^n$ by $\le$, i.e., for $x,y\in\mathbb{R}^n$, $x\leq y$ if and only if $x_i \le y_i$ for all $i\in \{1,\ldots,n\}$.  
A (bounded) \emph{interval} of $\mathbb{R}^n$ is a set of form $\{z : \ulx \le z \le \olx\}=:[\ulx,\olx]$ for some endpoints $\ulx,\olx\in\R^n$, $\ulx\leq \olx$.
Let $\IR^n$ denote the set of all intervals on $\R^n$. 
We also use the notation $[x]\in\IR^n$ to denote an interval when its endpoints are not relevant or implicitly understood to be $\ulx$ and $\olx$. For every two vectors $v,w\in \R^n$ and every $i\in \{1,\ldots,n\}$, we define the vector $v_{\{i:w\}}\in \mathbb{R}^n$ by $ \left(v_{\{i:w\}}\right)_j = \begin{cases}
    v_j & j\ne i\\
    w_j & j = i.
    \end{cases}$.
For a function $f:\R^n\to\R^m$ and a set $\calX\subseteq\R^n$, define the set-valued extension $f(\calX) = \{f(x) : x\in\calX\}$.
For two vectors $x,y\in\R^n$, let $(x,y)\in\R^{2n}$ denote their concatenation. 

\section{Interval Analysis} \label{sec:ianalysis}

\subsection{Interval Arithmetic}
Interval analysis extends operations and functions to intervals~\cite{LJ-MK-OD-EW:01}. For example, if we know that some number $a\in[\ula,\ola]$, and $b\in[\ulb,\olb]$, it is easy to see that the sum $(a+b)\in[\ula+\ulb, \ola+\olb]$. 

\begin{definition}[Inclusion Function~\cite{LJ-MK-OD-EW:01}]
    \label{def:if}
    Given a function $f:\R^n\to\R^m$, the interval function $[f]:\IR^n\to\IR^m$ is called an
    \begin{enumerate}
        \item \textit{inclusion function} for $f$ if, for every $[x]\in\IR^n$, $f([x]) \subseteq [f]([x])$;
        \item \textit{$[y]$-localized inclusion function} for $f$ if for every $[x]\subseteq [y]$, we have $f([x]) \subseteq [f]([x])$.
    \end{enumerate}
    Moreover, an inclusion function $[f]$ for $f$ is
    \begin{enumerate}\setcounter{enumi}{2}
        \item \textit{monotone} if $[x] \subseteq [y]$ implies that $ [f]([x])\subseteq[f]([y])$. 
        \item \textit{tight} if, for every $[x]$, $[f]([x])$ is the smallest interval containing $f([x])$.
    \end{enumerate}
\end{definition}

 In the next Theorem, we provide a closed-form expression for the tight inclusion function. 

\begin{theorem}[Uniqueness and Monotonicity of the Tight Inclusion Function]
    \label{thm:uniquemontif}
    Given a function $f:\R^n\to\R^m$, the tight inclusion function can be characterized as
    \begin{align*}
        [f]([x]) = \bigg[\inf_{x\in[x]} f(x), \sup_{x\in[x]} f(x)\bigg],
    \end{align*}
    where the $\inf$ and $\sup$ are taken element-wise, which is unique and monotone.
\end{theorem}
For some common functions, the tight inclusion function is easily defined. For example, if a function is monotonic, the tight inclusion function is simply the interval created by the function evaluated at its endpoints. 

More generally, the tight inclusion function can be alternatively computed using the fact that on a closed and bounded interval, a continuous function will achieve its maximal and minimal values at either the endpoints or a critical value in the interval. For any bounded interval, the tight inclusion function can be evaluated by taking the maximum and minimum of the function on each critical point within and the endpoints of the input interval. Tight inclusion functions for some elementary functions such as $\sin$ and $\cos$ can be defined in this manner (see Table~\ref{tab:tightif}).
However, when considering general functions, finding the tight inclusion function is often not computationally viable. The following theorem shows a more computational approach, by chaining known inclusion functions.

\begin{theorem}[Natural Inclusion Functions]
    \label{thm:nif}
    Given a function $f:\R^n\to\R^m$ defined by a composition of functions with known monotone inclusion functions, i.e., $f = e_\ell \circ e_{\ell-1} \circ \cdots \circ e_1$, an inclusion function for $f$ is formed by replacing each composite function with its inclusion function, i.e. $[f] = [e_\ell] \circ [e_{\ell-1}] \circ \cdots \circ [e_1]$ and is called the natural inclusion function. Additionally, if each of the $[e_j]$ are monotone inclusion functions, the natural inclusion function is also monotone.
\end{theorem}

\begin{figure}
    \centering
    \includegraphics[width=\columnwidth]{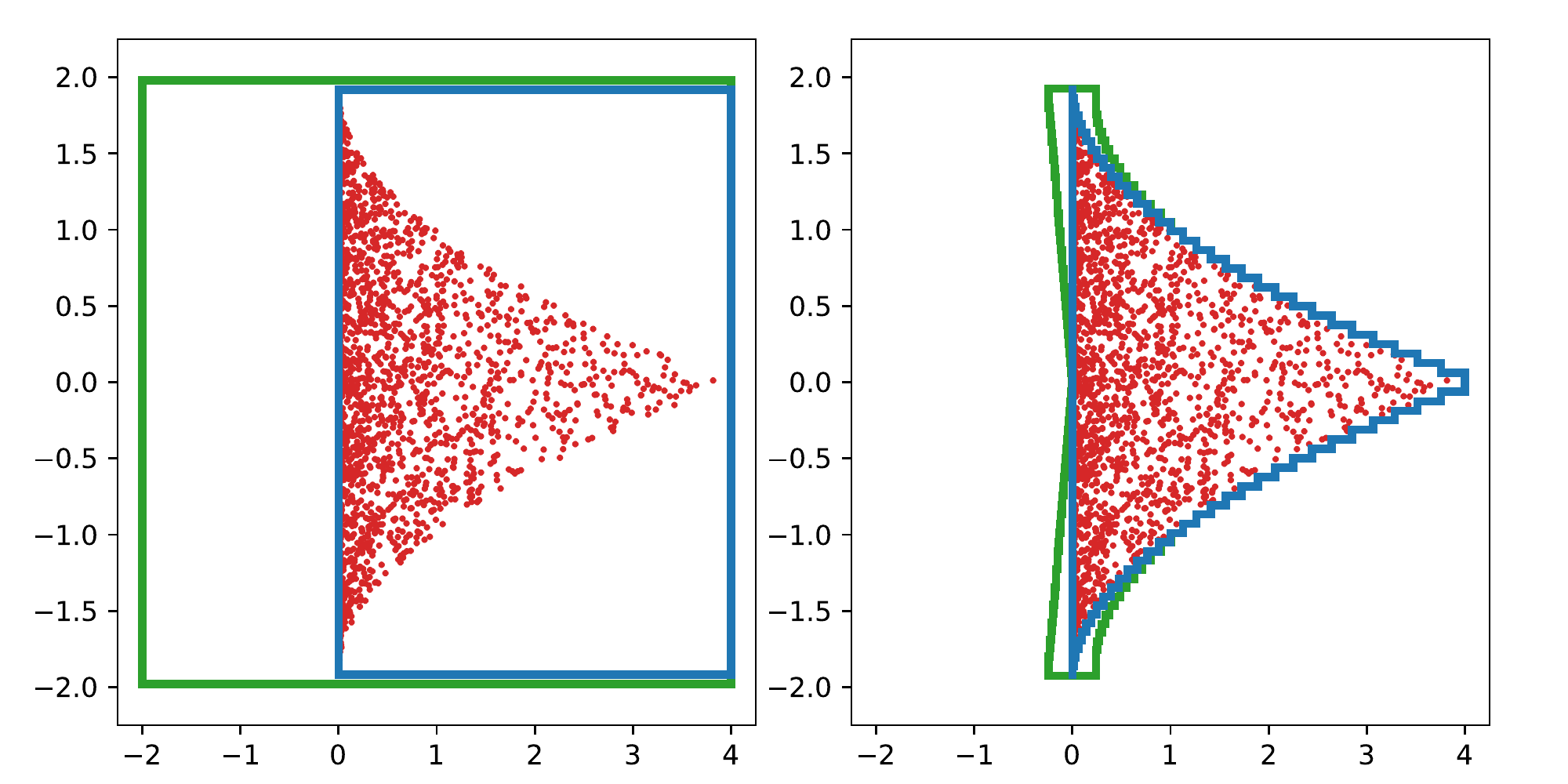}
    \caption{\textbf{Left:} \texttt{npinterval} is used to generate interval approximations for a function $f$ using two different natural inclusion functions. {\color{tab:blue} Blue:} $f(x_1,x_2) = [(x_1 + x_2)^2, 4\sin((x_1 - x_2)/4)]^T $
    {\color{tab:green} Green:} $f(x_1,x_2) = [x_2^2 + 2x_1x_2 + x_2^2, 4\sin(x_1/4)\cos(x_2/4) - 4\cos(x_1/4)\sin(x_2/4)]^T$. The approximations are generated using the initial set $[-1,1]\times[-1,1]$, and 2000 uniformly sampled ouptuts are shown in {\color{tab:red} red}. \textbf{Right:} The same function is analyzed, with the same two natural inclusion functions, but the initial set is partitioned into 1024 uniform sections, and the union of the interval approximations are shown.}
    \label{fig:interval_example1}
\end{figure}

Note that two different decompositions of the function $f$ can lead to two different natural inclusion functions for $f$. Thus, the natural inclusion function is not guaranteed to be the tight inclusion function. For example, consider the function
\[(x + 1)^2 = x^2 + 2x + 1,\]
on the interval $[-1,1]$. With the natural inclusion function for the first expression (LHS), the output interval is $([-1,1]+1)^2 = [0,4]$. With the natural inclusion function for the second expression (RHS), the output interval is $[-1,1]^2 + 2*[-1,1] + 1 = [0,1] + [-2,2] + 1 = [-1,4]$. Figure~\ref{fig:interval_example1} demonstrates this phenomenon in further detail.

\subsection{Automated Interval Analysis using \texttt{npinterval}}
The main contribution of this paper is to introduce the open source
\texttt{npinterval} package, an extension of \texttt{numpy} to allow native support for interval arithmetic. 
\texttt{npinterval} defines a new \texttt{interval} data-type, internally represented as a tuple of two \texttt{double}s, $[a] = (a.l,a.u)$.  The \texttt{interval} type is fully implemented in C, and therefore the standard operations from Table~\ref{tab:tightif} are all compiled into efficient machine-code executed as needed at runtime.
Additionally, since \texttt{interval} is implemented as a \texttt{dtype} in \texttt{numpy}, all of \texttt{numpy}'s prominent features, including $n$-dimensional arrays, fast matrix multiplication, and vectorization, are available to use.
In particular, each function from Table~\ref{tab:tightif} is registered as a \texttt{numpy} universal function, allowing for quick, element-wise operation over an $n$-dimensional array. 
While there are existing interval arithmetic toolboxes, none plug directly into \texttt{numpy}, opting instead to rewrite every operation in Python. While these packages do support the same standard operations from Table~\ref{tab:tightif}, they lose the flexibility and utility of \texttt{numpy}, as well as the efficiency of compiled C code.

\section{Interval Reachability of Neural Network Controlled Systems}

One application of interval analysis is in reachability analysis of neural network controlled systems. This section revisits and extends the framework considered in \cite{SJ-AH-SC:23}.

\subsection{Problem Statement}
Consider a dynamical system of the following form
\begin{align} \label{eq:nlsys}
    \dot{x} = f(x,u,w),
\end{align}
where $x\in\R^n$ is the system state, $u\in\R^p$ is the control input, $w\in\calW\subseteq\R^q$ is a unknown disturbance input in a compact set $\calW$, and $f:\R^n\times\R^p\times\R^q\to\R^n$ is a parameterized vector field. 
We assume that a feedback control policy for the system~\eqref{eq:nlsys} is given by a $k$-layer fully connected feed-forward neural network $N:\R^n\rightarrow\R^p$ as follows:
\begin{gather}\label{eq:NN}
    \begin{gathered}
        \xi^{(i)} = \sigma^{(i-1)} \left(W^{(i-1)}\xi^{(i-1)} + b^{(i-1)}\right),\, i=1,\ldots k \\
        \xi^{(0)} = x,\quad N(x) = W^{(k)} \xi^{(k)} + b^{(k)}
    \end{gathered}
\end{gather}
where $m_i$ is the number of neurons in the $i$-th layer, $W^{(i)} \in \R^{m_i\times m_{i-1}}$ is the weight matrix on the $i$-th layer, $b^{(i)}\in\R^{m_i}$ is the bias vector on the $i$-th layer, $\xi^{(i)} \in \R^{m_i}$ is the $i$-th layer hidden variable and $\sigma_i$ is the activation function for the $i$-th layer. Thus, we consider the closed-loop system 
\begin{align}
    \label{eq:clsys}
    \dot{x} = f(x,N(x),w) = f^c(x,w).
\end{align}
Given an initial time $t_0$, an initial state $x_0$, and a piecewise continuous mapping $\bfw:\R\to\calW$, denote the trajectory of the system for any $t\geq t_0$ as $\phi_{f^c}(t, t_0, x_0, \bfw)$. Given an initial set $\calX_0$, we denote the reachable set of $f^c$ at some $t\geq t_0$:
\begin{gather}
    \calR_{f}(t,t_0,\calX,\calW) = \left\{
    \begin{aligned}
    \phi&_{f^c}(t, t_0, x_0, \bfw),\,\forall x_0\in\calX_0,\\&w:\R\rightarrow\calW \text{ piecewise cont.}
    \end{aligned}
    \right\}
\end{gather}
One can use the reachable set of the system to verify safety specifications, e.g. by ensuring an empty intersection with unsafe states for all time. However, in general, computing the reachable set exactly is not computationally tractable---instead, approaches typically compute an over-approximation $\ol{\calR}_{f^{c}}(t,t_0,\calX,\calW)\supseteq\calR_{f^{c}}(t,t_0,\calX,\calW)$. The main challenge addressed in this section is to develop an approach for providing tight over-approximations of reachable sets while remaining computationally tractable for runtime computation.

\subsection{Open-loop System Interval Reachability}
Previously, \cite{SJ-AH-SC:23} consider a known decomposition function of the open-loop system. 
The interval analysis framework from Section~\ref{sec:ianalysis} allows us to extend this theory to remove the need to define a decomposition function \textit{a~priori}.
\begin{assumption}
    \label{assum:mifsys}
    For the dynamical system \eqref{eq:nlsys}, there exists a known monotone inclusion function $[f]$ for $f$. 
\end{assumption}
Using Theorem~\ref{thm:nif}, this assumption reduces to knowing a particular form $f=e_\ell\circ\cdots\circ e_1$ with known monotone inclusion functions for each $e_j$, thus removing the need to manually define a decomposition function for $f$.

The open-loop embedding function is defined by:
\begin{align}
\begin{gathered}
    \ul{\sfF}_i([x],[u],[w]) = \ulf_i([\ulx,\olx_{\{i:\ulx\}}],[u],[w]), \\
    \ol{\sfF}_i([x],[u],[w]) = \olf_i([\ulx_{\{i:\olx\}},\olx],[u],[w]),
\end{gathered}
\end{align}
for every $i\in\{1,\ldots,n\}$, where $[f]=[\ulf,\olf]$ is the monotone inclusion function of $f$, and $\ul{\sfF},\ol{\sfF} : \IR^n\times\IR^p\times\IR^q\to\R^n$. Using this embedding function, the following embedding dynamics can be defined
\begin{align}
    \label{eq:olembdyn}
    \begin{bmatrix}
        \dot{x} \\ \dot{\widehat{x}}
    \end{bmatrix} = 
    \begin{bmatrix}
        \ul{\sfF}([x,\widehat{x}],[u],[w]) \\ \ol{\sfF}([x,\widehat{x}],[u],[w])
    \end{bmatrix},
\end{align}
where $(x,\widehat{x})\in\R^{2n}$, $x\leq \widehat{x}$.
\begin{theorem}[Open-Loop System Interval Reachability]
    \label{thm:olsysreach}
    Let $t\mapsto (\ulx(t),\olx(t))$ denote the trajectory of the system \eqref{eq:olembdyn} with initial condition $(\ulx_0,\olx_0)$, under interval control mapping $[\bfu]: \R\to \IR^p$ and interval disturbance mapping $[\bfw]:\R\to\IR^q$. Let $t\mapsto x(t)$ denote the trajectory of the system \eqref{eq:nlsys} with initial condition $x_0$, under control $\bfu:\R\to\R^p$ and disturbance $\bfw:\R\to\calW$. If $x_0\in[\ulx_0,\olx_0]$, $\bfu(t)\in[\bfu](t)$, and $\bfw(t)\in[\bfw](t)$ for every $t\ge t_0$, then
    \[
        x(t)\in [\ulx(t),\olx(t)],\quad \mbox{ for every }t \ge t_0. 
    \]
\end{theorem}

\subsection{Interconnected Closed-Loop System Interval Reachability}
While Theorem~\ref{thm:olsysreach} provides a method of over-approximating the system $\eqref{eq:clsys}$ using global bounds on the control input, it disregards all interactions between the neural network controller and the dynamical system.

\begin{assumption}
    \label{assum:mifnn}
    For the neural network~\eqref{eq:NN}, there exists a known monotone inclusion function $[N]$ for $N$.
\end{assumption}
For example, one can use CROWN~\cite{HZ-etal:18} to obtain these bounds. Using CROWN, given an interval $[y]$, we can obtain affine upper and lower bounds of the following form
\begin{align}
    \label{eq:CROWN}
    \ulC_{[y]} x + \uld_{[y]} \le N(x) \le \olC_{[y]} x + \old_{[y]},
\end{align}
valid for any $x\in[y]$, which can be used to create the following monotone $[y]$-localized inclusion function $[N]_{[y]} = [\ulN_{[y]},\olN_{[y]}]$, with
\begin{align}
    \label{eq:CROWNmif}
    \begin{gathered}
        \ulN_{[y]}([x]) = \ulC_{[y]}^+ \ulx + \ulC_{[y]}^-\olx + \uld_{[y]}, \\
        \olN_{[y]}([x]) = \olC_{[y]}^+ \olx + \olC_{[y]}^-\ulx + \old_{[y]}, \\
    \end{gathered}
\end{align}
valid for any $[x]\subseteq[y]$. One can then construct the following ``hybrid'' closed-loop embedding function by interconnecting the open-loop embedding function with the monotone inclusion function for the neural network as follows
\begin{align}\label{eq:FC}
    \begin{gathered}
        \ul{\sfF}^c_i([x],[w]) = \ulf_i([\ulx,\olx_{\{i:\ulx\}}], [N]_{[x]}([\ulx,\olx_{\{i:\ulx\}}]), [w]), \\
        \ol{\sfF}^c_i([x],[w]) = \olf_i([\ulx_{\{i:\olx\}},\olx], [N]_{[x]}([\ulx_{\{i:\olx\}},\olx]), [w]), \\
    \end{gathered}
\end{align}
for every $i\in\{1,\ldots,n\}$, and $\ul{\sfF}^c,\ol{\sfF}^c:\IR^n\times\IR^q\to\R^n$. Using this embedding function, the following embedding dynamics can be defined
\begin{align}
    \label{eq:clembdyn}
    \begin{bmatrix}
        \dot{x} \\ \dot{\widehat{x}}
    \end{bmatrix} = 
    \begin{bmatrix}
        \ul{\sfF}^c([x,\widehat{x}],[w]) \\ \ol{\sfF}^c([x,\widehat{x}],[w])
    \end{bmatrix},
\end{align}
where $(x,\widehat{x})\in\R^{2n}$, $x\leq \widehat{x}$.

\begin{theorem}[Closed-Loop System Interval Reachability]
    \label{thm:clsysreach}
    Let $t\mapsto (\ulx(t),\olx(t))$ denote the trajectory of the system \eqref{eq:clembdyn} with initial condition $(\ulx_0,\olx_0)$, under interval disturbance mapping $[\bfw]:\R\to\IR^q$. Let $t\mapsto x(t)$ denote the trajectory of the closed-loop system \eqref{eq:clsys} with initial condition $x_0$, under disturbance $\bfw:\R\to\calW$. If $x_0\in[\ulx_0,\olx_0]$ and $\bfw(t)\in[\bfw](t)$ for every $t\ge t_0$, then
    \[
        x(t)\in [\ulx(t),\olx(t)],\quad\mbox{ for every } t\ge t_0.
    \]
    
\end{theorem}

\subsection{Experiments}
\paragraph*{Vehicle Model}
Consider the nonlinear dynamics of a vehicle adopted from~\cite{PP-FA-BdAN-AdlF:17}:
\begin{xalignat}{2} 
    \label{eq:vehicle1}
     \dot{p_x} &= v \cos(\phi + \beta(u_2)) & \dot{\phi} &=\frac{v}{\ell_r}\sin(\beta(u_2))\\
    \label{eq:vehicle2}
     \dot{p_y} &= v \sin(\phi + \beta(u_2)) & \dot{v} &= u_1.
\end{xalignat} 
where $[p_x,p_y]^{\top}\in \R^2$ is the displacement of the center of mass, $\phi \in [-\pi,\pi)$ is the heading angle in the plane, and $v\in \R_{\geq 0}$ is the speed of the center of mass. Control input $u_1$ is the applied force, input $u_2$ is the angle of the front wheels, and $\beta(u_2) = \mathrm{arctan}\left(\frac{\ell_f}{\ell_f+\ell_r}\tan(u_2)\right)$ is the slip slide angle. Let $x=[p_x,p_y,\phi,v]^\top$ and $u=[u_1,u_2]^\top$. We use the neural network controller ($4\times 100\times 100\times 2$ ReLU) defined in \cite{SJ-AH-SC:23}, which is applied at evenly spaced intervals of $0.25$ seconds apart. This neural network is trained to mimic an MPC that stabilizes the vehicle to the origin while avoiding a circular obstacle centered at $(4,4)$ with a radius of $2$. 

\begin{figure}
    \centering
    \includegraphics[width=0.6\columnwidth]{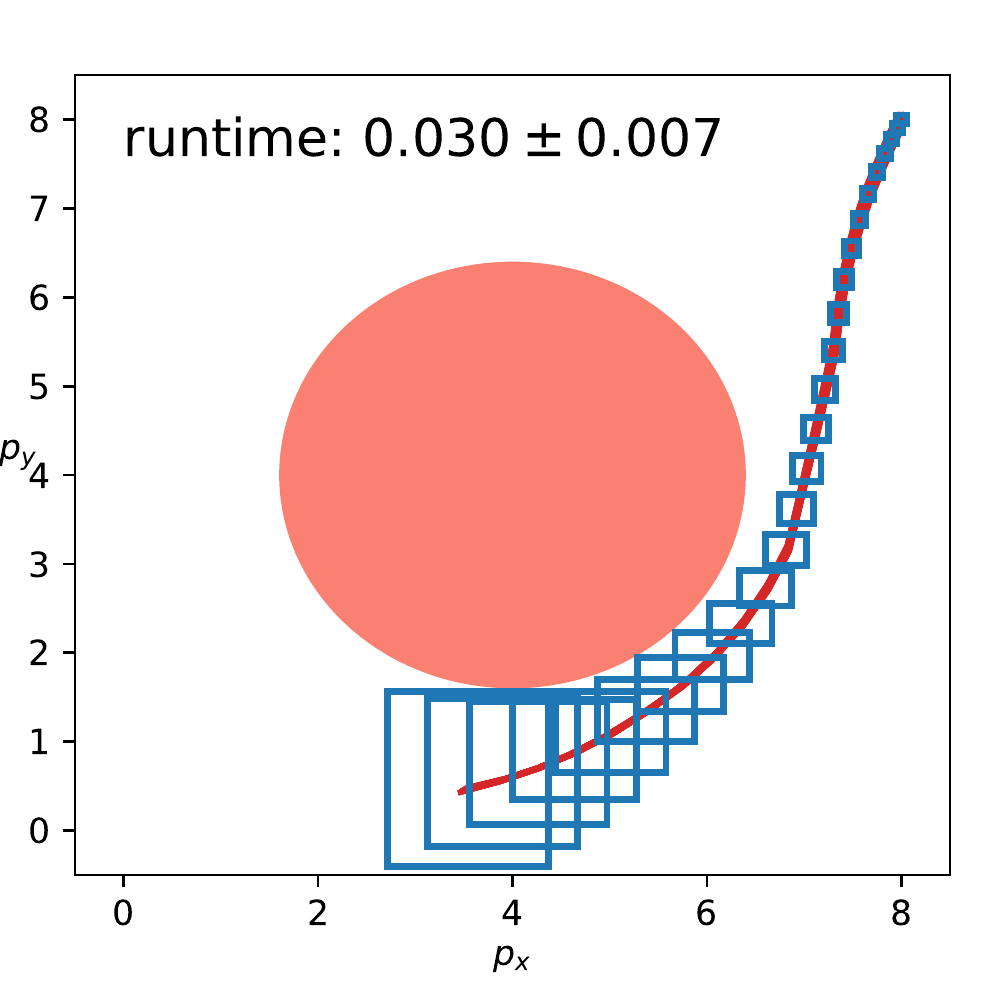}
    \caption{The over-approximated reachable set of the nonlinear vehicle model in the $p_x$-$p_y$ coordinates are shown in blue for the initial set $[7.95,8.05]^2\times[-\frac{2\pi}{3}-0.005, -\frac{2\pi}{3}+0.005]\times[1.995,2.005]$ over the time interval $[0,1.25]$. 100 true trajectories of the system are shown in red, and the average runtime and standard deviation over 100 runs is shown.}
    \label{fig:vehicle}
\end{figure}

The natural inclusion function is constructed using \texttt{npinterval} with Table~\ref{tab:tightif}, and the monotone inclusion function \eqref{eq:CROWNmif} is computed using \texttt{autoLiRPA} \cite{autolirpa}.
The closed-loop embedding function \eqref{eq:clembdyn} is then used to over-approximate the reachable set of the system using Theorem~\ref{thm:clsysreach}. The dynamics are simulated using Euler integration with a step size of $0.05$. The results are shown in Figure~\ref{fig:vehicle}.

\section{Conclusion}
In this paper, we introduced a framework for interval analysis implemented directly in \texttt{numpy}, called \texttt{npinterval}. The framework provides an automatic way to generate provide interval bounds on the output of a general class of functions. We use this framework to formally verify the output of nonlinear neural network controlled systems.
In the future, \texttt{npinterval} will be updated to account for floating point error, as well as to support a wider array of standard inclusion functions.

\section*{Acknowledgements}

This work was supported in part by the National Science Foundation under grant \#2219755 and by the Ford Motor Company.

\bibliography{refs}
\bibliographystyle{wfvml2023}

\newpage
\appendix
\onecolumn
\section{Table of Tight Inclusion Functions and Operations Implemented in \texttt{npinterval}}

\begin{table}[htp]
\caption{Tight inclusion functions and operations supported by \texttt{npinterval}.}
\label{tab:tightif}
\vskip 0.15in
\begin{center}
\begin{small}
\begin{sc}
\begin{tabular}{cc}
\toprule
Function/Operation & Tight inclusion function/Implementation \\
\midrule
$[a] + c$ & $[\ula + c, \ola + c]$\\
$c*[a]$ & $\begin{cases}
    \left[c\ula, c\ola \right]& c\ge 0 \\
    [c\ola,c\ula] & c < 0
\end{cases}$ \\
$[a] + [b]$ & $[\ula + \ulb, \ola + \olb]$ \\
$[a] - [b]$ & $[\ula - \olb, \ola - \ulb]$ \\
$[a] * [b]$ & $[\min \{\ula\ulb,\ula\olb,\ola\ulb,\ola\olb\}, \max\{\ula\ulb,\ula\olb,\ola\ulb,\ola\olb\}] $ \\
$1/[a]$ & $\begin{cases}
    \left[1/\ola, 1/\ula \right]& 0\notin [a] \\
    [-\infty,\infty] & 0\in [a]
\end{cases}$ \\
$[a]^n$ & $\begin{cases}
    [\ula^n, \ola^n] & n \text{ odd} \\
    [0,\max\{\ula^n, \ola^n\}] & n \text{ even, } 0\in[a] \\
    [\min\{\ula^n,\ola^n\}, \max\{\ula^n, \ola^n\}]  & n \text{ even, } 0\notin[a]
\end{cases}$ \\
$f([a]),\,f$ mon inc & $[f(\ula), f(\ola)]$ \\
$f([a]),\,f$ mon dec & $[f(\ola), f(\ula)]$ \\
Trigonometric & See \cref{eq:sinif,eq:tanif} \\
$\begin{gathered}
    [A]*[B],\\ \,[A]\in\IR^{n\times p},\,[B]\in\IR^{p\times m}
\end{gathered}$ & $[\cdot]_{i,j} = \sum_{k=1}^{p} [a_{i,k}] *[b_{k,j}]$ \\
\bottomrule
\end{tabular}
\end{sc}
\end{small}
\end{center}
\vskip -0.1in
\end{table}

The following is the tight inclusion function for $\sin$,
\begin{align} \label{eq:sinif}
    \sin([a]) = \begin{cases}
        [-1,1] & \calA_1 \\
        [\uls, \ols]& \calA_2 \land ((\ulc,\olc) \ge 0) \\
        [\ols, \uls]& \calA_2 \land ((\ulc,\olc) \le 0) \\
        [\min\{\uls,\ols\}, 1]& \calA_2 \land ((\ulc,\olc) \ge_{\text{SE}} 0) \\
        [-1, \max\{\uls,\ols\}]& \calA_2 \land ((\ulc,\olc) \le_{\text{SE}} 0) \\
        [-1, 1]& \calA_3 \land ((\ulc,\olc) \ge 0) \\
        [-1, 1]& \calA_3 \land ((\ulc,\olc) \le 0) \\
        [\min\{\uls,\ols\}, 1]& \calA_3 \land ((\ulc,\olc) \ge_{\text{SE}} 0) \\
        [-1, \max\{\uls,\ols\}]& \calA_3 \land ((\ulc,\olc) \le_{\text{SE}} 0) \\
    \end{cases},
\end{align}
where $\calA_{1} := (|[a]| > 2\pi)$, $\calA_2 := (\pi < |[a]| \le 2\pi)$, $\calA_3 = (0 < |[a]| \le \pi)$, $\uls := \sin(\ula)$, $\ols := \sin(\ola)$, and $\ulc := \cos(\ula)$, $\olc := \cos(\ola)$. 
Note that the tight inclusion function for $\cos$ can be defined as $\cos([a]) = \sin([a] + \pi/2)$. 
\begin{align} \label{eq:tanif}
    \tan([a]) = \begin{cases}
        [\tan(\ula), \tan(\ola)] & \forall k\in\N,\,\frac{\pi}{2} + \pi k \notin[a] \\
        [-\infty, \infty] & \exists k\in\N,\, \frac{\pi}{2} + \pi k \in[a] 
    \end{cases}
\end{align}

So far, in \texttt{npinterval}, the implemented monotonic functions include $\exp, \log, \arctan,$ and $\operatorname{sqrt}$, and will be extended in the future. 

\newpage
\section{Proof of the main results}
\label{sec:proofs}
In this section, we define the southeast order $\le_{\mathrm{SE}}$ on $\R^n$ by $\begin{bmatrix}x\\y\end{bmatrix} \le_{\mathrm{SE}} \begin{bmatrix}\widehat{x}\\\widehat{y}\end{bmatrix}$ if and only if $x\le \widehat{x}$ and $\widehat{y}\le y$. 
\paragraph*{Proof of Theorem~\ref{thm:uniquemontif}}
Take $f:\R^n\to\R^m$, and fix $[x]\in\IR^n$. We need to show that the smallest interval containing $f([x])$ is $\left[\inf_{x\in[x]} f(x), \sup_{x\in[x]} f(x)\right]$.

Fix $i\in\{1,\ldots,n\}$. For contradiction, assume that the smallest interval containing the set $f_i([x])$ is not $\left[\inf_{x\in[x]} f_i(x), \sup_{x\in[x]} f_i(x)\right]$. Then, there are two (non-exclusive) cases:

\textbf{Case 1:} There exists $a > \inf_{x\in[x]} f_i(x)$ such that $f_i(x)\ge a$ for every $x\in[x]$.
However, by the definition of $\inf$, there exists $x'\in[x]$ such that $\inf_{x\in[x]} f_i(x) < f_i(x') < a$, which is a contradiction.

\textbf{Case 2:} There exists $b < \sup_{x\in[x]} f_i(x)$ such that $f_i(x)\le b$ for every $x\in[x]$.
However, by the definition of $\sup$, there exists $x'\in[x]$ such that $b < f_i(x') < \sup_{x\in[x]} f_i(x)$, which is a contradiction.

Thus, the smallest interval containing the set $f_i([x])$ is $\left[\inf_{x\in[x]} f_i(x), \sup_{x\in[x]} f_i(x)\right]$. This is true for every $i$, completing the proof.

\paragraph{Proof of Theorem~\ref{thm:nif}}
We prove the theorem by induction.

\textbf{Base Case:} $m=1$

Since $[e_1]$ is an inclusion function for $e_1$, we have that $e_1 \subseteq [e_1]([x])$ for any interval $[x]$. Moreover, since $[e_1]$ is a monotone inclusion function, we have $[x]\subseteq[y]\implies [e_1]([x])\subseteq[e_1]([y])$.

\textbf{Inductive Step:} $[e_{m-1}]\circ\cdots\circ[e_1]$ monotone inclusion function $\implies$ $[e_m]\circ[e_{m-1}]\circ\cdots\circ[e_1]$ monotone inclusion function

Assume that $[e_{m-1}]\circ\cdots\circ[e_1]$ is a monotone inclusion function for $e_{m-1} \circ\cdots\circ e_1$. For every interval $[x]$, we have that $e_m \circ e_{m-1} \circ\cdots\circ e_1([x]) = e_m(e_{m-1} \circ\cdots\circ e_1([x])) \subseteq e_m([e_{m-1}] \circ\cdots\circ [e_1]([x])) \subseteq [e_m]([e_{m-1}] \circ\cdots\circ [e_1]([x]))$, which implies that $[e_m]\circ [e_{m-1}] \circ\cdots\circ [e_1]$ is an inclusion function for $e_m \circ e_{m-1} \circ\cdots\circ e_1$.

Moreover, since $[e_{m-1}]\circ\cdots\circ[e_1]$ is a monotone inclusion function, we have that for any two intervals $[x]\subseteq[y]$, $[e_{m-1}]\circ\cdots\circ[e_1]([x]) \subseteq [e_{m-1}]\circ\cdots\circ[e_1]([y])$. Since $[e_m]$ is monotone, we also have $[e_m]([e_{m-1}]\circ\cdots\circ[e_1]([x])) \subseteq [e_m]([e_{m-1}]\circ\cdots\circ[e_1]([y]))$, implying that $[e_m]\circ [e_{m-1}]\circ\cdots\circ[e_1]$ is a monotone inclusion function.

This completes the proof.

\paragraph{Proof of Theorem~\ref{thm:olsysreach}}

We define the function $d$ for the open-loop dynamics $f$~\eqref{eq:nlsys} as follows:
 \begin{align}\label{eq:tight-openloop}
     d_i(x,\widehat{x},u,\widehat{u},w,\widehat{w}) &= \begin{cases}
     \min_{z\in [x,\widehat{x}],\xi\in [w,\widehat{w}]\atop z_i=x_i, \eta \in [u,\widehat{u}]} f_i(z,\eta,\xi), & x\le \widehat{x},u\le \widehat{u},w\le\widehat{w}\\
     \max_{z\in [\widehat{x},x],\xi\in [w,\widehat{w}]\atop z_i=\widehat{x}_i,\eta \in [u,\widehat{u}]} f_i(z,\eta,\xi), &  \widehat{x}\le x , \widehat{u}\le u, \widehat{w}\le w, 
     \end{cases}
 \end{align}
 and we consider the following dynamical system on $\R^{2n}$: \begin{align}\label{eq:openloop-embedding}
     \frac{d}{dt}\begin{bmatrix}
         x\\\widehat{x} 
     \end{bmatrix} = \begin{bmatrix}  d(x,\widehat{x},u,\widehat{u},w,\widehat{w})\\d(\widehat{x},x,\widehat{u},u,\widehat{w},w)
     \end{bmatrix}
 \end{align}
The trajectory of the embedding system~\eqref{eq:openloop-embedding} with the control input $\left[\begin{smallmatrix}u\\ \widehat{u} \end{smallmatrix}\right] = \left[\begin{smallmatrix}\underline{u}\\ \overline{u} \end{smallmatrix}\right]$ and disturbance $\left[\begin{smallmatrix}w\\ \widehat{w} \end{smallmatrix}\right] = \left[\begin{smallmatrix}\underline{w}\\ \overline{w} \end{smallmatrix}\right]$ starting from $\left[\begin{smallmatrix}\underline{x}_0\\ \overline{x}_0\end{smallmatrix}\right]$  is denoted by $t\mapsto \left[\begin{smallmatrix}\underline{x}^{o}(t)\\ \overline{x}^{o}(t)\end{smallmatrix}\right]$. Note that, by~\citep[Theorem 1]{MA-MD-SC:21}, we have $x(t) \in [\underline{x}^{o}(t),\overline{x}^{o}(t)]$, for every $t\in \R_{\ge 0}$. Moreover, for every $i\in\{1,\ldots,n\}$, we get
 \begin{align}\label{eq:inequality-tight}
    d_i(x,\widehat{x},u,\widehat{u},w,\widehat{w}) & = \min_{z\in [x,\widehat{x}],\xi\in [w,\widehat{w}]\atop z_i=x_i, \eta \in [u,\widehat{u}]} f_i(z,\eta,\xi) \ge    \underline{\sfF}_i([x,\widehat{x}],[u], [w]), 
 \end{align}
where the first equality holds by definition of $d$ and second inequality holds by Theorem~\eqref{thm:uniquemontif}. Similarly, one can show that $d_i(\widehat{x},x,\widehat{u},u,\widehat{w},w) \le \overline{\sfF}_i([x,\widehat{x}],[u],[w])$, for every $i\in \{1,\ldots,n\}$. This implies that $\left[\begin{smallmatrix}\underline{\sfF}_i([x,\widehat{x}], [u],[w])\\\overline{\sfF}_i([x,\widehat{x}],[u],[w])\end{smallmatrix}\right] \le_{\mathrm{SE}} \left[\begin{smallmatrix}d(x,\widehat{x},u,\widehat{u},w,\widehat{w})\\ d(\widehat{x},x,\widehat{u},u,\widehat{w},w)\end{smallmatrix}\right]$, for every $x\le \widehat{x}$ and every $w\le \widehat{w}$ and every $u\le \widehat{u}$. Note that, by~\citep[Theorem 1]{MA-MD-SC:21}, the vector field $\left[\begin{smallmatrix}d(x,\widehat{x},u,\widehat{u},w,\widehat{w})\\ d(\widehat{x},x,\widehat{u},u,\widehat{w},w)\end{smallmatrix}\right]$ is monotone with respect to the southeast order $\le_{\mathrm{SE}}$ on $\R^{2n}$.  Now, we can use~\citep[Theorem 3.8.1]{ANM-LH-DL:08}, to deduce that $\left[\begin{smallmatrix}\underline{x}(t)\\\overline{x}(t)\end{smallmatrix}\right] \le_{\mathrm{SE}} \left[\begin{smallmatrix}\underline{x}^{o}(t)\\\overline{x}^o(t)\end{smallmatrix}\right]$, for every $t\in \R_{\ge 0}$. This implies that $[\underline{x}^{o}(t),\overline{x}^{o}(t)]\subseteq [\underline{x}(t),\overline{x}(t)]$. On the other hand, by~\citep[Theorem 1 and Theorem 2]{MA-MD-SC:21}, we know that $x(t)\in [\underline{x}^{o}(t),\overline{x}^{o}(t)]$, for every $t\in \R_{\ge 0}$. This lead to $x(t)\in [\underline{x}(t),\overline{x}(t)]$, for every $t\in \R_{\ge 0}$.

\paragraph{Proof of Theorem~\ref{thm:clsysreach}}

We define the function $d^{c}$ for the closed-loop system $f^c$~\eqref{eq:clsys} as follows:
 \begin{align}\label{eq:tight-closedloop}
     d_i^{c}(x,\widehat{x},w,\widehat{w}) &= \begin{cases}
     \min_{z\in [x,\widehat{x}],\xi\in [w,\widehat{w}]\atop z_i=x_i} f^c_i(z,N(z),\xi), & x\le \widehat{x},w\le\widehat{w}\\
     \max_{z\in [\widehat{x},x],\xi\in [w,\widehat{w}]\atop z_i=\widehat{x}_i} f^c_i(z,N(z),\xi), &  \widehat{x}\le x , \widehat{w}\le w.
     \end{cases}
 \end{align}
 and we consider the following dynamical system on $\R^{2n}$: \begin{align}\label{eq:closedloop-embedding}
     \frac{d}{dt}\begin{bmatrix}
         x\\\widehat{x} 
     \end{bmatrix} = \begin{bmatrix}
         d^{c}(x,\widehat{x},w,\widehat{w})\\d^{c}(\widehat{x},x,\widehat{w},w)
     \end{bmatrix}
 \end{align}
The trajectory of the embedding system~\eqref{eq:closedloop-embedding} with disturbance $\left[\begin{smallmatrix}w\\ \widehat{w} \end{smallmatrix}\right] = \left[\begin{smallmatrix}\underline{w}\\ \overline{w} \end{smallmatrix}\right]$ starting from $\left[\begin{smallmatrix}\underline{x}_0\\ \overline{x}_0\end{smallmatrix}\right]$  is denoted by $t\mapsto \left[\begin{smallmatrix}\underline{x}^{c}(t)\\ \overline{x}^{c}(t)\end{smallmatrix}\right]$. Note that, by~\citep[Theorem 1 and Theorem 2]{MA-MD-SC:21}, we have $x(t) \in [\underline{x}^{c}(t),\overline{x}^{c}(t)]$, for every $t\in \R_{\ge 0}$. 

Let $i\in \{1,\ldots,n\}$ and $y \in [x,\widehat{x}]$ be such that $y_i=x_i$. Note that $[ \underline{N}_{[x,\widehat{x}]} , \overline{N}_{[x,\widehat{x}]}]$ is an monotone inclusion function for $N$ on $[x,\widehat{x}]$. Moreover, $y\in [x,\widehat{x}_{[i:x]}]\subseteq [x,\widehat{x}]$ and thus
\begin{align}\label{eq:keyinequality}
     \underline{N}_{[x,\widehat{x}]}(x,\widehat{x}_{[i:x]}) \le N(y) \le \overline{N}_{[x,\widehat{x}]}(x,\widehat{x}_{[i:x]}).  
 \end{align}
Therefore, for every $i\in\{1,\ldots,n\}$, we get
 \begin{align}\label{eq:inequality-tight}
    d_i^{c}(x,\widehat{x},w,\widehat{w}) & = \min_{z\in [x,\widehat{x}],\xi\in [w,\widehat{w}]\atop z_i=x_i} f^c_i(z,N(z),\xi) \ge \min_{z\in [x,\widehat{x}],\xi\in [w,\widehat{w}], z_i=x_i\atop u\in [\underline{N}_{[x,\widehat{x}]}(x,\widehat{x}_{[i:x]}),\overline{N}_{[x,\widehat{x}]}(x,\widehat{x}_{[i:x]})]} f^c_i(z,u,\xi) \nonumber \\ & \ge \underline{f}^c_i([x,\widehat{x}],[N]_{[x]}({[x,\widehat{x}_{\{i:x\}}]}) , [w]) = \underline{\sfF}^c_i([x,\widehat{x}],[w]), 
 \end{align}
where the first equality holds by definition of $d^{c}$, the second inequality holds by equation~\eqref{eq:keyinequality}, the third inequality holds by Theorem~\eqref{thm:uniquemontif}, and the fourth inequality holds by definition of $\underline{\sfF}^c$ in equation~\eqref{thm:uniquemontif}. Similarly, one can show that $d_i^{c}(\widehat{x},x,\widehat{w},w) \le \overline{\sfF}^c_i([x,\widehat{x}],[w])$, for every $i\in \{1,\ldots,n\}$. This implies that $\left[\begin{smallmatrix}\underline{\sfF}^c_i([x,\widehat{x}],[w])\\\overline{\sfF}^c_i([x,\widehat{x}],[w])\end{smallmatrix}\right] \le_{\mathrm{SE}} \left[\begin{smallmatrix}d^{c}(x,\widehat{x},w,\widehat{w})\\ d^{c}(\widehat{x},x,\widehat{w},w)\end{smallmatrix}\right]$, for every $x\le \widehat{x}$ and every $w\le \widehat{w}$. Note that, by~\citep[Theorem 1]{MA-MD-SC:21}, the vector field $\left[\begin{smallmatrix}d^{c}(x,\widehat{x},w,\widehat{w})\\ d^{c}(\widehat{x},x,\widehat{w},w)\end{smallmatrix}\right]$ is monotone with respect to the southeast order $\le_{\mathrm{SE}}$ on $\R^{2n}$.  Now, we can use~\citep[Theorem 3.8.1]{ANM-LH-DL:08}, to deduce that $\left[\begin{smallmatrix}\underline{x}(t)\\\overline{x}(t)\end{smallmatrix}\right] \le_{\mathrm{SE}}\left[\begin{smallmatrix}\underline{x}^{c}(t)\\\overline{x}^c(t)\end{smallmatrix}\right]$, for every $t\in \R_{\ge 0}$. This implies that $[\underline{x}^{c}(t),\overline{x}^{c}(t)]\subseteq [\underline{x}(t),\overline{x}(t)]$. On the other hand, by~\citep[Theorem 1 and Theorem 2]{MA-MD-SC:21}, we know that $x(t)\in [\underline{x}^{c}(t),\overline{x}^{c}(t)]$, for every $t\in \R_{\ge 0}$. This lead to $x(t)\in [\underline{x}(t),\overline{x}(t)]$, for every $t\in \R_{\ge 0}$.

\end{document}